# A Comparative Examination of Network and Contract-Based Blockchain Storage Solutions for Decentralized Applications

Lipeng He*

Department of Combinatorics and Optimization, University of Waterloo, 200 University Avenue West, Waterloo, ON, Canada N2L 3G1

```
*Corresponding author: lipeng.he@uwaterloo.ca
```

**Abstract.** Decentralized applications (DApps), which are innovative blockchain-powered software systems designed to serve as the fundamental building blocks for the next generation of Internet services, have witnessed exponential growth in recent years. This paper thoroughly compares and analyzes two blockchain-based decentralized storage networks (DSNs), which are crucial foundations for DApp and blockchain ecosystems. The study examines their respective mechanisms for data persistence, strategies for enforcing data retention, and token economics. In addition to delving into technical details, the suitability of each storage solution for decentralized application development is assessed, taking into consideration network performance, storage costs, and existing use cases. By evaluating these factors, the paper aims to provide insights into the effectiveness of these technologies in supporting the desirable properties of truly decentralized blockchain applications. In conclusion, the findings of this research are discussed and synthesized, offering valuable perspectives on the capabilities of these technologies. It sheds light on their potential to facilitate the development of DApps and provides an understanding of the ongoing trends in blockchain development.

## 1    Introduction

Blockchain technology has rapidly evolved beyond its initial use case of being a platform for digital currencies such as Bitcoin [1]. Today, it finds widespread application in a myriad of domains [2], one of which is decentralized storage systems. This paper aims to delve deep into decentralized storage solutions, spotlighting the potential they hold for decentralized applications. Specifically, it offers a comparative analysis of two leading blockchain-based storage solutions – Arweave [3], a network-based decentralized storage network, and Filecoin [4], a contract-based DSN. Decentralized storage solutions have demonstrated their advantages over centralized cloud storage, eliminating single points of failure [5] and providing a higher level of security and availability at a considerably lower cost [6]. Arweave and Filecoin have established themselves as front-runners in this domain, each with a unique approach to achieving decentralized storage. This paper breaks down and compares their technical underpinnings, such as their data persistence mechanisms and data retention enforcement methods. Moreover, it inspects the token economics of the native cryptocurrency used in each network, analyzing their resistance to inflation. The suitability of Arweave and Filecoin in serving the unique needs of decentralized applications is presented in Section 6, where the comparative evaluation will include network performance, storage cost, and a review of existing use cases. It is essential to underline that the comparison is not intended to establish the superiority of one over the other but rather to shed light on the areas of strength and weakness in each system to guide blockchain developers, researchers, and users in their decision-making processes. This comprehensive comparison aims to provide insights into the capabilities and limitations of both Arweave and Filecoin as blockchain-based storage solutions for DApps, stimulating future advancements and innovation in this exciting and dynamic field. In the following sections, a brief background on blockchain technology, DApps, and decentralized storage networks is presented. Thereafter, a discussion on the specifics of Arweave and Filecoin, along with an in-depth qualitative analysis, is covered. Finally, a discussion on the implications of the findings is included, along with conclusions.

## 2    Background

### 2.1    Blockchain technology

A blockchain is a decentralized and distributed ledger that operates on a peer-to-peer network. It consists of a continuously growing chain of blocks, each containing transaction records and securely linked together through cryptographic hashes [7]. Each block includes the hash value of the previous block, a timestamp, and a limited amount of transaction data [6]. Consensus among the network's majority of nodes is required for blocks to be accepted and added to the blockchain. Popular consensus mechanisms include Proof-of-Work (PoW) and Proof-of-Stake (PoS) [8]. The ledger is replicated across all participating nodes, with each node storing a copy of the entire blockchain. This design ensures practical immutability, transparency, cryptographic security, and accessibility of the stored data. The emergence of Bitcoin, a cryptocurrency utilizing distributed ledger technology (DLT), brought significant attention to blockchain technology in recent years. However, Bitcoin primarily focuses on electronic cash transactions and has been criticized by some economists as a potential Ponzi scam [9]. As the advantages of blockchain technology became apparent, researchers and industries recognized its potential beyond finance [10]. This led to the development of Ethereum, a general-purpose blockchain that introduced the concepts of smart contracts and decentralized applications [11]. Smart contracts aim to enable the authentication and autonomous execution of legal agreements while minimizing reliance on trusted intermediaries [12]. In Ethereum, smart contracts are typically written in Solidity, a Turing-complete programming language. During deployment, the contracts are compiled into executable bytecode and included in a transaction submitted to the blockchain network [11]. By inheriting the benefits of blockchain technology and extending its applications to computational processes, smart contracts enable blockchain technology to be utilized in various non-financial sectors, including supply chain management, social networking platforms, Internet of Things, and peer-to-peer cloud storage.

## 2.2 Decentralized applications

Smart contracts opened up new opportunities for blockchain technology to be used in a wide range of ways. Utilizing smart contracts, a novel form of a blockchain-empowered software system called decentralized applications can coordinate the possession and transfer of digital assets in the form of tokens and achieve many of the same things that traditional systems normally do in a fully decentralized manner [10]. In a previous survey [9] of decentralized applications, researchers have identified four main characteristics of DApps: 1) Publicly available source code, 2) Quantifiable credits and transactions among users, 3) Transparent through decentralized consensus, 4) No central or single point of failure similar to a fully decentralized peer-to-peer network [13]. A token is a commonly used terminology for the representation of assets in a blockchain system [14]. There are two types of tokens, fungible and non-fungible. Where fungible tokens typically refer to cryptocurrencies, and non-fungible tokens (NFTs) can be used to symbolize virtually everything. Depending on the design and purpose of the DApp, NFTs could represent photographs, videos, audio files, and other forms of intellectual property. NFTs serve to authenticate the uniqueness of digital assets, establishing them as non-interchangeable [15]. One of the most formidable challenges in the development of decentralized applications is the storage placement of NFT content data [16]. The entity of a digital asset, which is essentially the content data of an NFT, often takes the form of a high-resolution image, a full-size video or audio file. Due to the append-only and immutable-through-replication characteristics of a blockchain, storing large files directly on the blockchain is often expensive and highly inefficient [13]. Centralized cloud storage services such as Amazon Web Service (AWS) and Simple Storage Service (S3) carry inherent design limitations due to the risk associated with having a single point of failure. These models are also susceptible to threats such as Denial of Service (DoS) attacks, data intrusions, and outbound data threats [5, 13]. To support and retain the desirable characteristics of decentralized applications and the value of NFTs (i.e., economically efficient transactions, no single point of failure), many solutions have been developed that combine the use of off-chain distributed storage networks with only references (often in the form of hash values) being kept on-chain [16]. In the rest of the paper, two state-of-the-art blockchain-based solutions with distinct architectural designs currently employed and commonly used are examined.

## 2.3    Interplanetary file system

The Interplanetary File System (IPFS) [17] is a peer-to-peer data storage and sharing network created by Protocol Labs that consists of a suite of subprotocols. It allows files to be distributed through a content-addressed protocol in a decentralized way. In IPFS, content is identified based on its cryptographic hash and not its location [18]. A Merkle direct acyclic graph (DAG) of blocks identifiable by their content identifiers (CIDs) is created for files and directories. Peers or nodes in the IPFS network store and exchange the hashes and provide the service of locating and retrieving files for users through a routing method powered by distributed hash tables (DHTs). The design of IPFS enables efficient, censorship-resistant, and immutable storage of data [19]. However, there is no built-in incentivization mechanism in IPFS that guarantees the redundancy and availability of files stored on the network. Content is available only for as long as it is maintained by at least one online and discoverable network participant [20]. Although files are not actively replicated, they can be "pinned" to prevent their deletion by nodes. Nevertheless, storing and constantly pinning files could be complicated and costly, making it unsuitable for some use cases [24].

## 2.4    Blockchain-based decentralized storage networks (DSNs)

As previously described, blockchain technology has been used across all kinds of domains. Among its various applications, decentralized storage systems stand out as one of its most significant and impactful [2]. A decentralized storage network coordinates storage providers without single trusted parties to rent out their available hardware storage space for clients to store their data and offer file retrieval services in return for a profit. There are mainly two approaches to using blockchain for decentralized storage networks: 1) Blockchain as an incentive layer on top of a peer-to-peer storage network. In this approach, the blockchain layer manages the storing, verifying, updating and other operations about files on the network through smart contracts between clients and miners (a contract-based DSN). 2) Blockchain directly stores and maintains file data in blocks (a network-based DSN), and the content stored inherits all the security and availability properties from the underlying blockchain system [20]. The following sections discuss and compare one of the most widely used and representative solutions of each of the two approaches.

## 3    Filecoin: a contract-based decentralized storage network

Implemented on top of IPFS, Filecoin is a blockchain-based decentralized storage network and cryptocurrency that establishes an economic incentive system [19] for network nodes to provide data storage and retrieval services. Filecoin (FIL) is the native cryptocurrency of the blockchain that users and miners use for contract or protocol payments [13]. The Filecoin blockchain regulates two decentralized markets for storage providers and clients to interact with each other and form deals: the storage market and the retrieval market. Users pay for the storage and retrieval of their data. Those who provide storage space, known as storage miners, generate revenue by committing Filecoin sectors [22] to the protocol. Conversely, retrieval miners serve data through IPFS to the clients in return for a premium. Storage deals are valid for a certain period of time (from 6 to 18 months) [23]; after the lifetime of a contract has ended, renewal is required. Otherwise, storage providers lose all incentives to continue storing the files. By design, the objective of Filecoin is to establish a transparent, publicly verifiable, and incentive-driven DSN. Data integrity is achieved by requiring miners to provide collateral [2] for their service contracts. Data availability is ensured using two Proof-of-Storage consensus algorithms: Proof-of-Replication (PoRep) and Proof-of-Spacetime (PoSt) [6]. For audibility, the Filecoin blockchain stores and processes transaction records, responses to data integrity challenges, and a ledger for storage deals in the storage market [23]. Note that the orderbook of the retrieval market is stored off-chain in order to optimize read performance [19]. The blockchain employs a novel consensus algorithm called Expected Consensus (EC) for growing its chain of blocks [4]. Using EC, storage miners have the opportunity to mine blocks for the blockchain in accordance with the storage that they have committed to the network.

## 4     Arweave: A network-based decentralized storage network

Contrary to Filecoin, which provides only ephemeral storage services, Arweave is a decentralized storage network that employs a structure akin to a blockchain known as Blockweave [19] to facilitate a system for permanent on-chain data storage and incentive payments. The data structure of a blockweave is similar to a blockchain, with the only distinction being that a blockweave has a more complicated graph structure, while a blockchain is typically a singly linked list [3]. To address the inherent scalability issue of a blockchain requiring each node to store a complete replica of the entire chain, Arweave utilizes a mechanism similar to compact blocks (BIP-152) called Blockshadows. Blockshadowing works by separating transactions from the blocks themselves. Rather than transferring the entire block, only a minimal "blockshadow" is sent between nodes. This approach enables peers to reconstruct a complete block, thus reducing the amount of data that needs to be transmitted. A detailed discussion of the solution that Arweave uses to solve blockchain scalability issues regarding data storage is included in later sections. Arweave is designed to achieve an economically sustainable, immutable, timestamped [19], eternal ledger of history and knowledge [3]. Clients only pay a single upfront fee using Arweave's native cryptocurrency AR for storing their data, and after which files are stored on the Arweave network forever and become part of the consensus [22]. The consensus algorithm: Succinct Random Proofs of Access (SPoRA) is used by Arweave to incentivize miners to store as many files as possible by asking miners to recover previously stored data on the network randomly. This procedure is done through the inclusion of a "Recall Block" in the blockweave structure; details will be presented in the following section.

## 5    Technical Comparison of Arweave and Filecoin

### 5.1    Data persistence mechanism

At the core of Arweave's data persistence mechanism is the blockchain-like data structure Blockweave. In blockweave, every block is connected to two antecedent blocks by including their cryptographic hash pointers: one is the immediately preceding block in the "chain," and the other is a block from the earlier history of the blockchain, referred to as the Recall Block. A demonstration of recall blocks in a blockweave structure is shown in Figure 1.

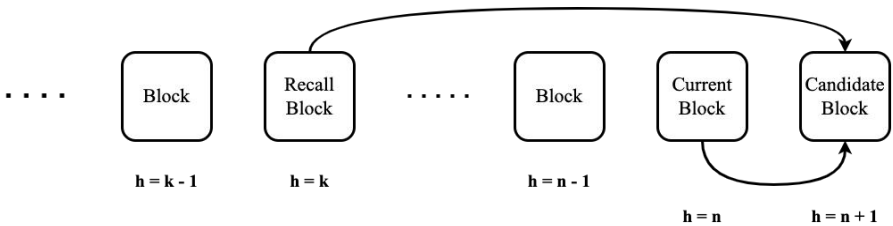

**Fig. 1.** Blockweave structure

Arweave integrates all data uploaded to its network on-chain. Each block in the Blockweave contains a transaction list, with each transaction comprising 0 to 10,485,760 bytes of arbitrary data and relevant metadata like the owner's RSA public key, tags, and digital signature. The Blockweave design guarantees that the entire block history is preserved and frequently accessed, ensuring the persistence and retrievability of data over time. Arweave's architecture provides the advantages and challenges typical of a blockchain system. In traditional blockchains, the entire content of a newly added block is broadcast to all network participants, irrespective of how many transactions in the block a node already stores. This distribution process is essential for consensus among nodes but could be too slow to prevent forks in the network if the data transferred is large [3]. To resolve the issue of balancing block size and the likelihood of a fork emerging during consensus, Arweave developed a unique solution called blockshadowing. Rather than distributing the entire block, a condensed version of the block, a blockshadow, is transmitted during consensus. This blockshadow only contains the list of transaction Merkle roots [19], a concise representation of the transaction data, not the transaction data themselves. This modification accelerates transaction propagation, making block distribution more efficient and cost-effective. For data retrieval, Arweave employs Wildfire, a system designed to expedite data requests within the network. Wildfire operates via a local ranking mechanism in each node that evaluates the responsiveness of its peers in answering requests and sending transactions. Inspired by BitTorrent's tit-for-tat protocol, each node maintains a scored list of peers, incentivizing nodes to be highly responsive to receive messages quickly. Nodes show a preference for engaging with higher-ranked peers, while underperforming nodes risk being disconnected from the network. This strategy optimizes resource utilization and reduces latency [19], aiming to deliver response speeds on par with traditional centralized storage providers. Conversely, Filecoin leverages blockchain technology in a markedly different manner. The Filecoin blockchain forms the basis for a decentralized storage marketplace where storage miners fulfill data storage and retrieval requests from clients, earning profits for rendering the contracted services. The actual data storage and retrieval processes are facilitated through IPFS. When a storage agreement is formed, a Filecoin sector is sealed, generating a Proof-of-Replication. The storage miner is then required to prove via a Proof-of-Spacetime, at thirty-second intervals (1 epoch) [23], that the committed data remains intact and appropriately stored. The Filecoin protocol incorporates all proofs, along with the CID of the data on IPFS and a list of transactions (or messages), in block headers (or blocks), storing them on the blockchain for transparency and verifiability. Since proofs are periodically generated and submitted to enhance network throughput and scalability, blocks in the Filecoin blockchain are assembled into Tipsets before being linked together. Therefore, the Filecoin blockchain is technically a linked list of Tipsets rather than blocks. The Expected Consensus algorithm elects multiple storage miners to mine new blocks in every epoch based on a Proof-of-Stake method [23], with a miner's stake equivalent to the amount of data storage they offer to the network. As shown in Table 1.

**Table 1.** Components overview of each DSN

| DSN | Data Storage | Data Retrieval | Consensus Algorithm |
|---|---|---|---|
| Arweave | Blockweave | Wildfire | Proof-of-Access / Proof-of-Work |
| Filecoin | IPFS | DHT | Proof-of-Storage (PoRep, PoSt) / Proof-of-Stake |

## 5.2 Data retention enforcement

Arweave protocol allows new blocks to be added to the blockweave through the classic Proof-of-Work consensus first introduced by Bitcoin, where miners must solve the computational puzzle [1] of finding the appropriate hash for a candidate block storing: 1) transactions and metadata of the current block, 2) the independent hash [3] of the previous block, 3) contents of the recall block. The inclusion of a reference to past data on the network as part of the construction and verification process of blocks enhances the original PoW consensus mechanism and hence was given a new name by Arweave called Proof-of-Access (PoA). The height $k$ of the recall block for the candidate block at height $n + 1$ can be determined based on the hash of the previous block and its height value $n - 1$. Although the algorithm is deterministic, the choice of recall block is unpredictable by the miners as it is based on previous block history [19]. The probability of a miner earning the block reward for adding a new valid block to the chain is proportionate to the average hashing power of all nodes in the network possessing the same recall block. Since nodes cannot pre-determine which block will be used as the recall block, they are incentivized to store as many previous blocks as possible to increase their probability of earning block rewards. This design results in a larger number of replications being stored by miners, increasing Arweave's network security and stability [3].

$$P(wins\ block\ reward) = P(has\ the\ recall\ block) \times P(solves\ PoW\ puzzle\ first) \quad (1)$$

While Arweave's PoA mechanism guarantees permanent storage and data access, it provides no incentives for miners to offer fast access to the stored files. Additionally, similar to PoW, PoA is considered energy-intensive and not environmentally friendly. As an attempt to solve these issues, after block height 633720, Arweave introduced a new consensus mechanism called Succinct Proofs of Random Access (SPoRA). SPoRA supersedes the classic Proof-of-Work (PoW) / Proof of Access (PoA) algorithms, bringing two main enhancements to the network: 1) Associate mining profitability with the rapidity of data access. SPoRA encourages efficient and swift data replication, which mitigates miners' attempts at retrieving data from the network on demand. 2) Reduced energy consumption by making it economically infeasible for miners to compensate for their lack of local data with computation. With the utilization of blockweave and SPoRA, Arweave provides clients with a sustainable, pay-once, store-forever storage solution. Compared to Arweave's lightweight design, Filecoin has a considerably more complex system. Filecoin employs zero-knowledge proofs (ZKPs) based cryptographic solutions for maintaining the execution of storage deals. In the Filecoin protocol, two Proof-of-Storage algorithms are used to ensure the honest behaviour of the storage providers: Proof-of-Replications (PoRep) and Proof-of-Spacetime. At the end of the settlement of a storage deal, the storage miner is required to generate a Proof-of-Replication for the network and the client themselves to verify that the desired data has been securely replicated to separate physical locations and that no nodes in the network keep the file twice [6]. Before committing the PoRep, a storage miner must seal a Filecoin sector containing the client's file data by computing its Merkle tree and encoding the sector. The miner proves the completion of sealing through a Succinct, Non-interactive Argument of Knowledge (SNARK) and locks in a certain amount of Filecoin currency as proof of their intention to store the sector. After the deal is settled and a PoRep proof is posted to the blockchain, the storage provider begins periodically generating Proofs-of-Spacetime to demonstrate to the network that the storage is maintained continuously as specified in the deal [2]. At any point in time, before the lifespan of a storage contract ends should the storage miner fail to generate valid PoSt proofs, the collateral will be lost [21]. Conversely, as a Filecoin sector or storage deal expires, the miner can collect the rewards along with the collaterals. The use of SNARK in both PoRep and PoSt is essential to supporting all Filecoin network participants to be able to verify the validity of proofs. These mechanisms work together to ensure the commitment of storage miners towards data retention in the Filecoin network, thereby providing a reliable, decentralized storage solution. By incorporating economic incentives and cryptographic proofs, Filecoin enforces data retention and maintains a high level of network integrity.

### 5.3 Token economics of AR and FIL

Arweave's native token AR gains its utility from acting as the sole medium of exchange for data storage on the network. AR has an initial circulating supply of 55 million, with an additional 11 million being gradually released as mining rewards [3]. Arweave's token economy offers three potential revenue avenues for nodes that contribute to growing the blockweave: rewards from inflation, immediate transaction-based earnings, and payments sourced from the endowment pool [14]. Different from traditional blockchains, transaction fees that clients paid for storing their data are not directly given to the miners as block mining awards, instead most of them goes into a storage endowment pool. The reward a miner receives for mining a new block can be calculated using the following formula:

$$R_{total} = R_{fees} + R_{inflation} + R_{endowment} \text{ [3]}, \tag{2}$$

where $R_{fees}$ is dependent on the size of the total data in transactions mined into the block, and $R_{inflation}$ is a pre-determined, decreasing value with respect to the height of blocks. Miners are only receiving funds from the endowment pool when $R_{fees} + R_{inflation}$ is not enough to cover the cost of storage maintenance [18]. Based on the principle of Moore's Law, the funds in the endowment pool are expected to generate interest over time. As the cost of data storage decreases, the interest accrued from the endowment pool should be sufficient to incentivize miners to store and serve data indefinitely. Filecoin's token economics revolves around its native cryptocurrency, FIL (Filecoin). FIL has a maximum total supply of 2 billion, but only 14,747,034 is available in circulation initially. The scarcity of FIL token is enforced through the following methods: 1) Open storage and retrieval markets. Storage miners earn FIL by providing storage to clients, while clients spend FIL to hire miners for storage and retrieval of data. 2) Collateral requirements in the consensus mechanism. If miners fail to provide PoSt, a portion of their locked collaterals will be slashed. 3) Token burning. In addition to the tokens burnt as penalties for consensus and storage faults, some network message fees are burnt for every storage deal. 4) Diminishing block mining rewards. 5) Mining reward vesting. The Filecoin network is designed to promote sustained participation and alignment with network goals. Specifically, miners receive 75% of their block rewards in a graduated manner over a period of 180 days, fostering long-term commitment. At the same time, to support miner's immediate operational needs and enhance profitability, the remaining 25% of the rewards are instantly accessible. These methods together act as deflationary counterweight to the inflation caused by the dynamic supply of FIL tokens.

## 6 Arweave and Filecoin for Decentralized Applications

### 6.1 Network performance comparison

Arweave's native protocol specifies the number of transactions stored in a block at a maximum of 1000. With a block time of 2 minutes, the expected throughput in terms of transactions per second (TPS) is eight on average. Network delay occurs when this limit is reached. However, a new network on top of Arweave called Bundlr is being actively developed and is said to be able to achieve over 50,000 TPS by bundling multiple transactions into one. Filecoin, on the other hand, has been reported to be much more challenging to produce reliable measurements of network performance. Although there are no official and accurate benchmarks available, a rough calculation done by one of the members of the Arweave team has prematurely concluded that as of Q3 2023, the Filecoin network has a TPS of 3,300 when assuming five blocks per Tipset.

## 6.2    Storage cost comparison

To support Arweave's goal of providing permanent storage for its clients, a special pricing mechanism was developed to achieve perpetual data storage. In short, transaction costs on the Arweave network are calculated by multiplying the size of the transaction data with the estimated cost of storing the data forever [3]. The cost estimation takes into account a 0.5% annual decrease in storage costs over the next 200 years and the total computational (hashing) power of the network. Since the purchasing power of AR tokens is independent of any fiat currency, the cost of storage on the Arweave network fluctuates depending on market demands. According to the statistics listed in Arweave's blockchain explorer, Arweave has an average cost of 0.858 AR per gigabytes of data stored, which is approximately 5,000 US dollars per TB. As offered as a reference by the Arweave protocol, a minimum of 200 years of retention is provided to the data stored, averaging the total cost of storing 1 TB of data at approximately 25 US dollars per year. Compared to Arweave, Filecoin has considerably more affordable pricing for short-term storage. Even with FIL's pricing constantly fluctuating in the storage marketplace, the cost of storing 1 TB of data for a year is less than 3 US dollars [13]. Storage price is charged on a yearly basis, regardless of FIL's exchange rate with fiat currencies. Failure to renew a storage contract could result in data loss. A summary of cost and performance findings for Arweave and Filecoin are presented in Table 2.

Table 2. Performance and cost overview of each DSN

| DSN | Transactions-Per-Second (TPS) | Bundled TPS | Cost Per TB of Data Store |
| --- | --- | --- | --- |
| Arweave | 8 | 50,000+ | 5000+ USD |
| Filecoin | 3,300 | - | 3 USD |

## 6.3    Existing use cases

While both Arweave and Filecoin are blockchain-based decentralized storage networks providing data storage and retrieval solutions, their use cases diverge due to their unique design features. Arweave is mainly used as a permanent, unalterable storage solution, and Filecoin is considered highly effective as a flexible, market-driven solution for large-scale data storage. Some of the existing use cases of Arweave include 1) Historical archiving. presented R-Archive as a tamper-proof and secure community archive and cultural resource that uses Blockweave technology to store and access records from anywhere in the world. 2) Decentralized websites. With the introduction of the Arweave network, a permanent and decentralized web built on top of Arweave's protocol called Permaweb was also conceptualized [3]. 3) Storing NFT content data with small sizes. Utilizing Arweave's technology to ensure complete on-chain storage of the artwork and tokens, InfiNFT [24] is a novel platform for minting valuable NFT. Filecoin, as an incentive layer on top of IPFS's storage network, opens up a wide variety of use cases such as 1) Large scale data backup, 2) Decentralized content delivery network (CDN), 3) Distributed applications, providing a decentralized storage backend for all kinds of decentralized applications.

# 7   Discussion

Arweave and Filecoin are two of the most widely used blockchain-based decentralized storage solutions for reliable digital asset storage in decentralized applications, taking first and second place in decentralized storage market share, according to industry research. Arweave's unique network-based design is the only solution in the market that offers sustainable permanent storage at a relatively low cost. Some researchers have deemed Arweave as the gold standard for NFT data storage. However, considering the significantly higher cost of storage and lower network throughput, Arweave's use cases are relatively limited when compared to Filecoin. As an incentive layer for IPFS and the contract-based storage and retrieval market, Filecoin aims to compete directly with traditional cloud storage providers, offering a decentralized and low-cost substitute for centralized services such as AWS. In addition to the potential but unlikely circumstances of a loss of data due to nodes going offline on the Filecoin network, researchers have found a growing trend in the centralization of Filecoin [23] due to its consensus design. Nevertheless, Filecoin still provides a lower barrier of entry for applications from all domains to transition from traditional centralized storage to decentralized storage solutions and accelerates the growth of blockchain ecosystems. Arweave, in its current state, might be more of an attraction for governments and high-worth individuals to store votes, identities, ownership proofs and other assets that would benefit from a non-ephemeral storage solution.

# 8   Conclusion

This paper presents a qualitative comparison of two leading blockchain-based, decentralized storage solutions that share common features, yet have different design objectives. It dissected the technical aspects, token economics, performance, cost, and use cases of each solution, highlighting both their strengths and weaknesses. Arweave and Filecoin both strive to deliver a transparent, secure, publicly verifiable, highly available, censorship-resistant, and decentralized storage system for data and information. Yet, they make specific trade-offs in their design to achieve distinct goals. Arweave's ambition lies in ensuring data perpetuity. Drawing from the innate characteristics of a blockchain system, Arweave optimizes essential components like the consensus mechanism and data replication method in order to attain scalability and compatibility ideal for data storage. In contrast, Filecoin seeks to disrupt the cloud storage industry, offering an alternative that delivers cost-effective, secure, and decentralized storage and retrieval services. When selecting a storage solution, developers of decentralized applications (DApps) need to carefully assess their data storage needs. Various factors such as user experience, costs, and the general goal of the DApp should be considered. This paper also shines a spotlight on the emerging blockchain-based storage network, Arweave. Despite its unique and captivating characteristics, research on Arweave is surprisingly sparse. This paper, therefore, hopes to inspire further studies into this intriguing project.